\documentclass{ws-procs9x6}

\newcommand{\ud}{\mathrm{d}}
\begin{document}
%%%%%%%%%%%%%%%%%%%%%%%%%%%%%%%%%%%%%%%%%%%%%%%%%%%%%%%%%%%%%%%%
\thispagestyle{empty}

\begin{center}
\begin{tabular}{p{130mm}}

\begin{center}
{\bf\Large
BBGKY DYNAMICS:} \\
\vspace{5mm}

{\bf\Large FROM LOCALIZATION TO PATTERN}\\
\vspace{5mm}

{\bf\Large FORMATION }\\

\vspace{1cm}

{\bf\Large Antonina N. Fedorova, Michael G. Zeitlin}\\

\vspace{1cm}

{\bf\it
IPME RAS, St.~Petersburg,
V.O. Bolshoj pr., 61, 199178, Russia}\\
{\bf\large\it e-mail: zeitlin@math.ipme.ru}\\
{\bf\large\it e-mail: anton@math.ipme.ru}\\
{\bf\large\it http://www.ipme.ru/zeitlin.html}\\
{\bf\large\it http://www.ipme.nw.ru/zeitlin.html}
\end{center}

\vspace{1cm}

\abstracts{
A fast and efficient numerical-analytical approach
is proposed for modeling complex behaviour
in the BBGKY--hierarchy  of kinetic equations.
Our calculations are based on                                          
variational and multiresolution 
approaches in the basis of            
polynomial tensor algebras of generalized coherent states/wavelets. 
We construct the representation for hierarchy of reduced distribution functions
via the multiscale
decomposition in
highly-localized eigenmodes.
Numerical modeling shows the creation of
various internal
structures from localized modes, which are related to localized or chaotic type
of behaviour and
the corresponding patterns (waveletons) formation. 
The localized pattern is a model for energy confinement
state (fusion) in plasma.
}

\vspace{20mm}

\begin{center}
{\large Presented at Workshop}\\
{\large "Progress in Nonequilibrium Greens Functions"} \\
{\large Dresden, Germany, August 19-23, 2002}
\end{center}
\end{tabular}
\end{center}
\newpage

%%%%%%%%%%%%%%%%%%%%%%%%%%%%%%%%%%%%%%%%%%%%%%%%%%%%%%%%%%%%%%%%
\title{BBGKY DYNAMICS: FROM LOCALIZATION TO PATTERN FORMATION
}
\author{Antonina~N.~FEDOROVA and Michael~G.~ZEITLIN}
\address{
IPME RAS, St.~Petersburg, 
V.O. Bolshoj pr., 61, 199178, Russia\\
E-mail: zeitlin@math.ipme.ru, anton@math.ipme.ru\\
http://www.ipme.ru/zeitlin.html\\
http://www.ipme.nw.ru/zeitlin.html}

\maketitle

\abstracts{
A fast and efficient numerical-analytical approach
is proposed for modeling complex behaviour
in the BBGKY--hierarchy  of kinetic equations.
Our calculations are based on                                          
variational and multiresolution 
approaches in the basis of            
polynomial tensor algebras of generalized coherent states/wavelets. 
We construct the representation for hierarchy of reduced distribution functions
via the multiscale
decomposition in
highly-localized eigenmodes.
Numerical modeling shows the creation of
various internal
structures from localized modes, which are related to localized or chaotic type
of behaviour and
the corresponding patterns (waveletons) formation. 
The localized pattern is a model for energy confinement
state (fusion) in plasma.
}

\section{Introduction}

Kinetic theory is an important part of general statistical physics
related to phenomena
which cannot be understood on the thermodynamic or fluid models level.
First of all we mean (local) fluctuations from equilibrium 
state and a lot of complex phenomena \cite{1}.
Also, it is well-known in plasma physics, e.g., that only a kinetic approach 
can describe Landau damping, 
intra-beam scattering, while Schottky noise and associated cooling 
techniques require understanding of the spectrum of local fluctuations
of the beam charge density \cite{2}.
In this paper we consider the applications of a new nu\-me\-ri\-cal/analytical 
technique based on wavelet analysis approach for 
calculations related to the description of complex behaviour 
in the framework of the general BBGKY--hierarchy \cite{1}. 
We restrict ourselves to rational/polynomial type of
nonlinearities (with respect to the set of all dynamical variables 
($n$-particle distribution functions in the case under consideration)) which 
allows us to use our results from 
Refs.~\cite{3}$^-$\cite{9}, which are based on the application of wavelet 
analysis technique and 
variational formulation of initial nonlinear (pseudodifferential) problems.

Wavelet analysis is a set of mathematical
methods which give us a possibility to work with well-localized bases in

functional spaces and provide maximum sparse forms  for the general 
type of operators (differential,
integral, pseudodifferential) in such bases. 
It provides the best possible rates of convergence and minimal complexity 
of algorithms inside and, 
as a result, saves CPU time and HDD space \cite{10}.
Our main goals are an attempt of classification and construction 
of possible nontrivial states
in the system under consideration.
First of all we are interested in the following states: localized, chaotic-like patterns,
localized (stable) patterns. 
We start from the corresponding definitions 
(at this stage these definitions have only
qualitative character).

1. By localized state (localized mode) 
we mean the corresponding (particular) solution of the system under 
consideration which is localized in maximally small region of the phase space.

2. By chaotic pattern we mean some solution (or asymptotics of solution) 
of the system under consideration
which has equidistribution of energy spectrum in a full domain of definition. 

3. By localized pattern (waveleton) 
we mean (asymptotically) stable solution localized in 
relatively small region of the whole phase space (or a domain of definition). 
In this case all energy is distributed during some time (sufficiently large) 
between few localized modes (from point 1) only. 
We believe, it is a good model for a plasma fusion (energy confinement) 
in the nearest future.
It is also obvious that such states are very important in many areas of statistical physics.

In all cases above, by the system under consideration we mean 
the full BBGKY--hierarchy or some cut-off of it.
Our construction of cut-off of the infinite system of equations 
is based on some criterion of convergence of the full solution.
This criterion
is based on a natural norm in a proper functional space, 
which takes into account (non-perturbatively) the underlying 
multiscale structure
of complex statistical dynamics.
According to our approach the choice of the underlying functional space is
 important for understanding the corresponding complex dynamics.
It is obvious that we need to fix accurately the space in which 
we construct the solutions, evaluate convergence etc. 
and in which the very complicated infinite set of operators acts
which appears in the BBGKY formulation.
We underline that many concrete 
features of the complex dynamics are 
related not only to the concrete 
form/class of operators/equations but depend also on the proper choice of 
function 
spaces, where operators act \cite{9}.
It should be noted that the class of smoothness of the functions 
under consideration 
plays a key role in the following (more details will be considered 
elsewhere).

In Sec. 2  the kinetic BBGKY--hierarchy is formulated and an important 
particular case is described.
In Sec. 3 we present the explicit analytical construction of solutions of
the hierarchy, which is based on tensor algebra
extensions of bases generated by the hidden 
multiresolution
structure and proper variational formulation leading 
to an algebraic parametrization of the solutions.
We give the explicit representation of the hierarchy of N-particle 
reduced distribution functions 
in the basis of
highly-localized generalized coherent (regarding the underlying affine group) 
states given by polynomial tensor algebra of wavelets, which 
takes into account
contributions from all underlying hidden multiscales,
from the coarsest scale of resolution to the finest one, to
provide full information about stochastic dynamical process.
So, our approach resembles Bogolyubov's and related approaches 
but we don't use any perturbation technique (like virial expansion)
or linearization procedures.
Numerical modeling as in general case as in particular cases of the  Vlasov-like equations 
shows the creation of
various internal (coherent)
structures from localized modes, which are related to stable (equilibrium) or unstable/chaotic type
of behaviour and the 
corresponding pattern (waveletons) formation.

\section{BBGKY-Hierarchy}

Let $M$ be the phase space of an ensemble of $N$ particles ($ {\rm dim}M=6N$)
with coordinates
$x_i=(q_i,p_i), \quad i=1,...,N, \quad
q_i=(q^1_i,q^2_i,q^3_i)\in R^3,\quad
p_i=(p^1_i,p^2_i,p^3_i)\in R^3,\quad
q=(q_1,\dots,q_N)\in R^{3N}$.
Individual and collective measures are: 
$
\mu_i=\ud x_i=\ud q_i\ud p_i,\quad \mu=\prod^N_{i=1}\mu_i
$.
The distribution function
$D_N(x_1,\dots,x_N;t)$
satisfies 
Liouville's equation  
of motion and the normalization constraint for an ensemble 
with the Hamiltonian $H_N$ :
\begin{eqnarray}
\frac{\partial D_N}{\partial t}=\{H_N,D_N\}\qquad 
\int D_N(x_1,\dots,x_N;t)\ud\mu=1.
\end{eqnarray}
Our constructions can be applied to the following general Hamiltonians:
\begin{eqnarray}
H_N=\sum^N_{i=1}\Big(\frac{p^2_i}{2m}+U_i(q)\Big)+
\sum_{1\leq i\leq j\leq N}U_{ij}(q_i,q_j),  
\end{eqnarray}
where the potentials 
$U_i(q)=U_i(q_1,\dots,q_N)$ and $U_{ij}(q_i,q_j)$
are restricted to rational functions of the coordinates.

Let $L_s$ and $L_{ij}$ be the Liouvillean operators (vector fields)
\begin{eqnarray}
L_s=\sum^s_{j=1}\Big(\frac{p_j}{m}\frac{\partial}{\partial q_j}-
\frac{\partial U_j}{\partial q}\frac{\partial}{\partial p_j}\Big)-
\sum_{1\leq i\leq j\leq s}L_{ij},
\end{eqnarray}
\begin{eqnarray}
L_{ij}=\frac{\partial U_{ij}}{\partial q_i}\frac{\partial}{\partial p_i}+
\frac{\partial U_{ij}}{\partial q_j}\frac{\partial}{\partial p_j}.
\end{eqnarray}
Let
$
F_N(x_1,\dots,x_N;t)=\sum_{S_N}D_N(x_1,\dots,x_N;t)
$
be the $N$-particle distribution function
($S_N$ is permutation group of N elements). 
Then we have the hierarchy of reduced distribution functions ($V$ is the
 volume) 
\begin{eqnarray}
F_s(x_1,\dots,x_s;t)=
V^s\int D_N(x_1,\dots,x_N;t)\prod_{s+1\leq i\leq N}\mu_i
\end{eqnarray}
After standard manipulations we arrive at the BBGKY--hierarchy \cite{1}:
\begin{eqnarray}
\frac{\partial F_s}{\partial t}+L_sF_s=\frac{1}{V}\int\ud\mu_{s+1}
\sum^s_{i=1}L_{i,s+1}F_{s+1}
\end{eqnarray}
It should be noted that we may apply our approach even to more 
general formulation (nonlinear) than
(6). 
For s=1,2 we have, from the general BBGKY--hierarchy:  
\begin{eqnarray}
\frac{\partial F_1(x_1;t)}{\partial t}+\frac{p_1}{m}
\frac{\partial}{\partial q_1}
F_1(x_1;t)
=\frac{1}{V}\int\ud x_2L_{12} F_2(x_1,x_2;t),\nonumber
\end{eqnarray}
\begin{eqnarray}
&&\frac{\partial F_2(x_1,x_2;t)}{\partial t}+\Big(\frac{p_1}{m}
\frac{\partial}{\partial q_1}+\frac{p_2}{m}
\frac{\partial}{\partial q_2}-L_{12}\Big)
 F_2(x_1,x_2;t)\\
&&=\frac{1}{V}\int\ud x_3(L_{13}+L_{23})F_3(x_1,x_2;x_3;t).\nonumber
\end{eqnarray}

In most cases, one is interested in a representation of the form 
$
F_k(x_1,\dots,x_k;t)=\prod^k_{i=1}F_1(x_i;t)+G_k(x_1,\dots,x_k;t),
$
where $G_k$ are correlators. Additional reductions often lead to 
simplifications, the simplest one, $G_k=0$, corresponding to the 
Vlasov approximation.
Such physically motivated ansatzes for $F_k$
formally replace the linear (in $F_k$) and pseudodifferential (in general case)
infinite system (6), (7) by
a finite-dimensional but nonlinear system with
 polynomial  nonlinearities (more exactly, multilinearities) \cite{10}.
Our key point in the following consideration is the proper 
generalization of the perturbative multiscale approach of Bogolyubov.

\section{Multiscale Analysis}

The infinite hierarchy of distribution functions satisfying system (6)
in the thermodynamical limit is:
\begin{eqnarray}
F=\{F_0,F_1(x_1;t),F_2(x_1,x_2;t),\dots,
F_N(x_1,\dots,x_N;t),\dots\},
\end{eqnarray}
where
$F_p(x_1,\dots, x_p;t)\in H^p$,
$H^0=R,\quad H^p=L^2(R^{6p})$ (or any different proper functional space), $F\in$
$H^\infty=H^0\oplus H^1\oplus\dots\oplus H^p\oplus\dots$
with the natural Fock space like norm 
(guaranteeing the positivity of the full measure):
\begin{eqnarray}
(F,F)=F^2_0+\sum_{i}\int F^2_i(x_1,\dots,x_i;t)\prod^i_{\ell=1}\mu_\ell.
\end{eqnarray}
First of all we consider $F=F(t)$ as a function of time only,
$F\in L^2(R)$, via
multiresolution decomposition which naturally and efficiently introduces 
the infinite sequence of the underlying hidden scales \cite{10}.
Because the affine
group  of translations and dilations 
generates multiresolution approach, this
method resembles the action of a microscope. We have the contribution to
the final result from each scale of resolution from the whole
infinite scale of spaces. 
We consider a multiresolution decomposition of $L^2(R)$ \cite{10}
(of course, we may consider any different and proper for some particular case functional space)
which is a sequence of increasing closed subspaces $V_j\in L^2(R)$ 
(subspaces for 
modes with fixed dilation value):
\begin{equation}
...V_{-2}\subset V_{-1}\subset V_0\subset V_{1}\subset V_{2}\subset ...
\end{equation}
The closed subspace
$V_j (j\in {\bf Z})$ corresponds to  the level $j$ of resolution, 
or to the scale j
and satisfies
the following properties:
let $W_j$ be the orthonormal complement of $V_j$ with respect to $V_{j+1}$: 
$
V_{j+1}=V_j\bigoplus W_j.
$
Then we have the following decomposition:
\begin{eqnarray}
\{F(t)\}=\bigoplus_{-\infty<j<\infty} W_j \qquad {\rm or} \qquad
\{F(t)\}=\overline{V_0\displaystyle\bigoplus^\infty_{j=0} W_j},
\end{eqnarray}
in case when $V_0$ is the coarsest scale of resolution.
The subgroup of translations generates a basis for the fixed scale number:
$
{\rm span}_{k\in Z}\{2^{j/2}\Psi(2^jt-k)\}=W_j.
$
The whole basis is generated by action of the full affine group:
\begin{eqnarray}
{\rm span}_{k\in Z, j\in Z}\{2^{j/2}\Psi(2^jt-k)\}=
{\rm span}_{k,j\in Z}\{\Psi_{j,k}\}
=\{F(t)\}.
\end{eqnarray}
Let the sequence $\{V_j^t\}, V_j^t\subset L^2(R)$ 
correspond to multiresolution analysis on the time axis, 
$\{V_j^{x_i}\}$ correspond to multiresolution analysis for coordinate $x_i$,
then
$
V_j^{n+1}=V^{x_1}_j\otimes\dots\otimes V^{x_n}_j\otimes  V^t_j
$
corresponds to the multiresolution analysis for 
the $n$-particle distribution function 
$F_n(x_1,\dots,x_n;t)$.
E.g., for $n=2$:$\qquad V^2_0=\{f:f(x_1,x_2)=$
{\setlength\arraycolsep{0mm}
$
\sum_{k_1,k_2}a_{k_1,k_2}\phi^2(x_1-k_1,x_2-k_2),\ 
a_{k_1,k_2}\in\ell^2(Z^2)\},
$}
where 
$
\phi^2(x_1,x_2)=\phi^1(x_1)\phi^2(x_2)=\phi^1\otimes\phi^2(x_1,x_2),
$
and $\phi^i(x_i)\equiv\phi(x_i)$ form a multiresolution basis corresponding to
$\{V_j^{x_i}\}$.
If $\{\phi^1(x_1-\ell)\},\ \ell\in Z$ form an orthonormal set, then 
$\phi^2(x_1-k_1, x_2-k_2)$ form an orthonormal basis for $V^2_0$.
So, the action of the affine group generates multiresolution representation of
$L^2(R^2)$. After introducing the detail spaces $W^2_j$, we have, e.g. 
$
V^2_1=V^2_0\oplus W^2_0.
$
Then the 
3-component basis for $W^2_0$ is generated by 
the translations of three functions 
\begin{eqnarray}
\Psi^2_1=\phi^1(x_1)\otimes\Psi^2(x_2),\ \Psi^2_2=\Psi^1(x_1)\otimes\phi^2(x_2), \ 
\Psi^2_3=\Psi^1(x_1)\otimes\Psi^2(x_2).\nonumber
\end{eqnarray}
Also, we may use the rectangle lattice of scales and one-dimensional wavelet
decomposition:
$$
f(x_1,x_2)=\sum_{i,\ell;j,k}\langle f,\Psi_{i,\ell}\otimes\Psi_{j,k}\rangle
\Psi_{j,\ell}\otimes\Psi_{j,k}(x_1,x_2),
$$
where the basis functions $\Psi_{i,\ell}\otimes\Psi_{j,k}$ depend on
two scales $2^{-i}$ and $2^{-j}$.
After constructing the multidimensional basis 
we may apply one of the variational procedures from
Refs.~\cite{3}$^-$\cite{9}.
We obtain our multiscale\-/mul\-ti\-re\-so\-lu\-ti\-on 
representations (formulae (17) below) 
via the variational wavelet approach for 
the following formal representation of the BBGKY system (6) 
(or its finite-dimensional nonlinear approximation for the 
$n$-particle distribution functions) 
with the
corresponding obvious constraints on 
the distribution functions.
 
Let $L$ be an arbitrary (non)li\-ne\-ar dif\-fe\-ren\-ti\-al\-/\-in\-teg\-ral operator 
 with matrix dimension $d$
(finite or infinite), 
which acts on some set of functions
from $L^2(\Omega^{\otimes^n})$:  
$\quad\Psi\equiv\Psi(t,x_1,x_2,\dots)=\Big(\Psi^1(t,x_1,x_2,\dots), \dots$,
$\Psi^d(t,x_1,x_2,\dots)\Big)$,
 $\quad x_i\in\Omega\subset{\bf R}^6$, $n$ is the number of particles:
\begin{equation}
L\Psi\equiv L(Q,t,x_i)\Psi(t,x_i)=0,
\end{equation}
where
\begin{eqnarray}
&&Q\equiv Q_{d_0,d_1,d_2,\dots}(t,x_1,x_2,\dots,\partial /\partial t,\partial /\partial x_1,
\partial /\partial x_2,\dots,\int \mu_k)=\nonumber\\
&&\sum_{i_0,i_1,i_2,\dots=1}^{d_0,d_1,d_2,\dots}
q_{i_0i_1i_2\dots}(t,x_1,x_2,\dots)
\Big(\frac{\partial}{\partial t}\Big)^{i_0}\Big(\frac{\partial}{\partial x_1}\Big)^{i_1}
\Big(\frac{\partial}{\partial x_2}\Big)^{i_2}\dots\int\mu_k. 
\end{eqnarray}
Let us consider now the $N$ mode approximation for the solution as 
the following ansatz:
\begin{equation}
\Psi^N(t,x_1,x_2,\dots)=\sum^N_{i_0,i_1,i_2,\dots=1}a_{i_0i_1i_2\dots} A_{i_0}\otimes 
B_{i_1}\otimes C_{i_2}\dots(t,x_1,x_2,\dots).
\end{equation}
We shall determine the expansion coefficients from the following conditions
(different related variational approaches are considered in 
\cite{3}$^-$\cite{9}):
\begin{equation}
\ell^N_{k_0,k_1,k_2,\dots}\equiv
\int(L\Psi^N)A_{k_0}(t)B_{k_1}(x_1)C_{k_2}(x_2)\ud t\ud x_1\ud x_2\dots=0.
\end{equation}
Thus, we have exactly $dN^n$ algebraical equations for  $dN^n$ unknowns 
$a_{i_0,i_1,\dots}$.
This variational approach reduces the initial problem 
to the problem of solution 
of functional equations at the first stage and 
some algebraical problems at the second.
 We consider the multiresolution expansion as the second main part of our 
construction. 
So, the solution is parametrized by the solutions of two sets of 
reduced algebraical
problems, one is linear or nonlinear
(depending on the structure of the operator $L$) and the rest are linear
problems related to the computation of the coefficients of the algebraic equations (16).
These coefficients can be found  by some wavelet methods
by using the
compactly supported wavelet basis functions for the expansions (15).
As a result the solution of the equations (6) has the 
following mul\-ti\-sca\-le or mul\-ti\-re\-so\-lu\-ti\-on decomposition via 
nonlinear high\--lo\-ca\-li\-zed eigenmodes 
%{\setlength\arraycolsep{0pt}
\begin{eqnarray}
&&F(t,x_1,x_2,\dots)=
\sum_{(i,j)\in Z^2}a_{ij}U^i\otimes V^j(t,x_1,x_2,\dots),\nonumber\\
&&V^j(t)=
V_N^{j,slow}(t)+\sum_{l\geq N}V^j_l(\omega_lt), \quad \omega_l\sim 2^l, \\
&&U^i(x_s)=
U_M^{i,slow}(x_s)+\sum_{m\geq M}U^i_m(k^{s}_mx_s), \quad k^{s}_m\sim 2^m,
 \nonumber
\end{eqnarray}
%}
which corresponds to the full multiresolution expansion in all underlying time/space 
scales.
The formulae (17) give the expansion into a slow part $\Psi_{N,M}^{slow}$
and fast oscillating parts for arbitrary $N, M$.  So, we may move
from the coarse scales of resolution to the 
finest ones for obtaining more detailed information about the dynamical process.
In this way one obtains contributions to the full solution
from each scale of resolution or each time/space scale or from each nonlinear eigenmode.
It should be noted that such representations 
give the best possible localization
properties in the corresponding (phase)space/time coordinates. 
Formulae (17) do not use perturbation
techniques or linearization procedures.
Numerical calculations are based on compactly supported
wavelets and related wavelet families \cite{10} and on evaluation of the 
accuracy on 
the level $N$ of the corresponding cut-off of the full system (6) 
regarding norm (9):
\begin{equation}
\|F^{N+1}-F^{N}\|\leq\varepsilon.
\end{equation}

To summarize, the key points are:

1. The ansatz-oriented choice of the (multi\-di\-men\-si\-o\-nal) ba\-sis
related to some po\-ly\-no\-mi\-al tensor algebra. 

2. The choice of a proper variational principle. A few 
pro\-je\-c\-ti\-on (Ga\-ler\-kin\--li\-ke) 
principles for constructing (weak) solutions are considered in
\cite{3}$^-$\cite{9}.
The advantages of formulations related to biorthogonal
(wavelet) decomposition should be noted. 

3. The choice of  basis functions in the scale spaces $W_j$. They 
correspond to highly-localized (nonlinear) oscillations/excitations, 
nontrivial local (stable) distributions/fluctuations,
etc. Besides fast convergence properties we note 
 the minimal complexity of all underlying calculations, 
especially by choosing wavelet
packets which minimize Shannon's entropy. 

4.  Operator  representations providing maximum sparse representations 
for arbitrary (pseudo) differential/ integral operators 
$\ud f/\ud x$, $\ud^n f/\ud x^n$, $\int T(x,y)f(y)\ud y)$, etc.~\cite{10}.

5. (Multi)linearization. Besides the variation approach we can consider 
also a different method
to deal with (polynomial) nonlinearities: 
para-products-like decompositions \cite{10,9}.

\section{Example: Vlasov equation}
As a particular case we consider the Vlasov approximations 
$F_2=F_1 F_1$, $F_3(x_1,x_2,x_3)=\sum_{S_3}F_1(x_i)F_2(x_j,x_k)$
in Eqs. (7),
which are important in plasma physics.  

This is a particular case of the general form (13) (for simplicity we consider
only one variable) 
\begin{equation}
Q(R,x) \Psi(x)=P(R,x)\Psi(x) \quad{\rm or}\quad
L\Psi\equiv L(R,x)\Psi(x)=0,
\end{equation}
where
$R\equiv R(x,\partial /\partial x, \Psi)$ is not more than a rational (operator) function.
We have the following 
representation
for 
the $N$ mode approximation for 
the solution of the Vlasov equation 
via 
expansion in some high-localized wavelet-like basis 
(other independent variables are considered analogously): 
\begin{equation}
\Psi^N(x)=\sum^N_{r=1}a^N_{r}\phi_r(x).
\end{equation}
We shall determine the expansion coefficients from the following 
variational conditions:
\begin{equation}
L^N_{k}\equiv\int(L\Psi^N)\phi_k(x)\ud x=0.
\end{equation}
We have exactly $dN$ algebraical equations for  $dN$ unknowns $a_{r}$.
\begin{figure}[htb]                                                                    
\centering                                                                             
\includegraphics*[width=60mm]{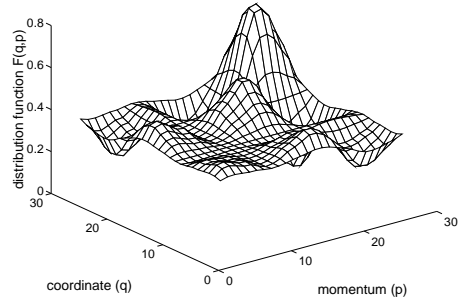}                                         
\caption{Localized mode contribution to distribution 
function.}                                                        
\end{figure}  
\begin{figure}[htb]
\centering
\includegraphics*[width=60mm]{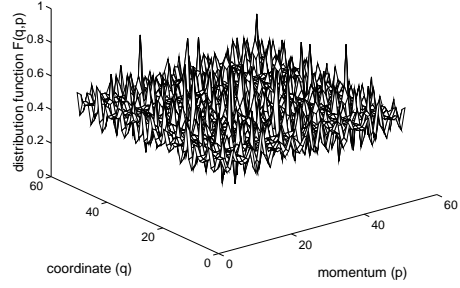}
\caption{Chaotic-like pattern.}
\end{figure} 
So, the variational approach reduces the initial problem (7) to the problem of the solution 
of functional equations at the first stage and some algebraical problems at the second
stage. 
As a result we have the following reduced algebraical system
of equations (RSAE) on the set of unknown coefficients $a_i^N$ of
the expansion (20):
\begin{eqnarray}
H(Q_{ij},a_i^N,\alpha_I)=M(P_{ij},a_i^N,\beta_J),
\end{eqnarray}
where the operators $H$ and $M$ are algebraizations of the RHS and LHS 
of the initial problem
(19).
$Q_{ij}$ ($P_{ij}$) are the coefficients of LHS (RHS) of the initial
system of differential equations (19) and, as consequence, are the coefficients
of the RSAE.
$I=(i_1,...,i_{q+2})$, $ J=(j_1,...,j_{p+1})$ are multiindexes
labeling $\alpha_I$ and $\beta_I$,  the other coefficients of (22) are
\begin{equation}
\beta_J=\{\beta_{j_1...j_{p+1}}\}=\int\prod_{1\leq j_k\leq p+1}\phi_{j_k},
\end{equation}
where $p$ is the degree of the polynomial operator $P$,
\begin{equation}
\alpha_I=\{\alpha_{i_1}...\alpha_{i_{q+2}}\}=\sum_{i_1,...,i_{q+2}}\int
\phi_{i_1}...\dot{\phi}_{i_s}...\phi_{i_{q+2}},
\end{equation}
and $q$ is the degree of the polynomial operator $Q$,
$i_\ell=(1,...,q+2)$, $\dot{\phi}_{i_s}=\ud\phi_{i_s}/\ud x$.

We may extend our approach to the case when there are additional
constraints on the set of dynamical variables and additional 
averaged terms also.
In these cases, by using the method of Lagrangian 
multipliers  
one again may apply the same 
approach, but
for the extended set of variables. As a result, one obtains the 
expanded system 
of algebraical equations,
analogous to the system (22). Then, one again can extract from its 
solution the coefficients 
of expansion (20).  
It should be noted that if one considers only the truncated expansion 
(22) with $N$ terms
one has a system of $N\times d$ algebraical equations
of degree $\ell=max\{p,q\}$
which coincides
with the degree of the initial system.
\begin{figure}[htb]
\centering
\includegraphics*[width=60mm]{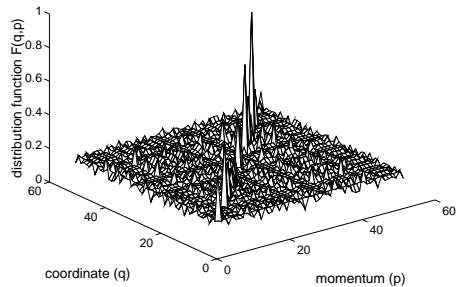}
\caption{Localized waveleton pattern.}
\end{figure} 

\section{Demonstration of typical patterns}
Formulae (15), (17), (20) provide, in principle, a fast convergent 
decomposition
for the general solutions of the systems (6), (7)
in terms of contributions from all underlying hidden internal scales.
Of course, we cannot guarantee that each concrete system (6), (7)
with fixed coefficients will have a priori a specific type of behaviour, 
either localized or chaotic. 
Instead, we can analyze if typical structures described by definitions
1-3 are present. 
To classify the qualitative behaviour we apply
standard methods from general control theory or really use the 
control \cite{9}. 
We will start from a priori unknown coefficients, the exact values of which 
will subsequently be recovered.
Roughly speaking, we will fix only class of nonlinearity 
(polynomial in our case)
which covers a broad variety of examples of systems (6), (7).

As a simple model we choose band-triangular 
non-sparse matrices $(a_{ij})$ from (17) or (20) in particular cases.
These matrices provide tensor structure of bases in (extended) phase space
and are generated by the roots of the reduced variational (Galerkin-like) 
systems (16) or (21), (22).
As a second step we need to restore the coefficients of (6), (7) from these 
matrices
by which we may classify the types of behaviour. 
We start with the localized mode, which is an ``elementary'' eigenfunction,
 Fig. 1, corresponding to def. 1, 
which was constructed as a tensor product of the two Daubechies functions. 
Fig. 2, corresponding to def. 2, 
presents the result of summation of series (17) up to value of the 
dilation/scale parameter equal to six
on the bases of symmlets \cite{10} with the corresponding matrix 
elements equal to one.  The size of matrix
is 512x512 and as a result we provide modeling for one-particle 
distribution function corresponding to
standard Vlasov like-system (7) with 
$F_2=F_1^2$. So, different possible  distributions of  the root values 
of the generical 
algebraical system (22) provide  qualitatively different types of behaviour. 
The above choice
provides us by chaotic-like equidistribution distribution.
But, if we consider a band-like structure of matrix $(a_{ij})$ 
with the band along the main diagonal with 
 finite size ($\ll 512$) and values, e.g. five, while the other 
values are equal to one, we obtain
 localization in a fixed finite area of the full phase space, 
i.e. almost all energy of the system 
is concentrated in this small volume. This corresponds to definition 3 and 
is shown in Fig. 3, constructed by means of Daubechies-based wavelet packets. 
Depending on the type of 
solution,  such localization may be present during the whole time evolution
(asymptotically-stable) or up to the needed value from time scale (e.g. enough 
for plasma fusion/confinement).

Now we discuss how to solve the inverse/synthesis problem or how to restore
the coefficients of the initial systems (6), (7).
Let 
\begin{equation}
L^0(Q^0)\Psi^0=0
\end{equation}
be the system (13) with the fixed coefficients $Q^0$. 
The corresponding solution $\Psi^0$
is represented by formulae (15), (17) or (20), 
which are parametrized by roots of reduced algebraic
system (22) and constructed by some choice of the tensor product 
bases from part 1.
The proper counterpart of the system (25) with prescribed behaviour 
$\Psi^u$, 
corresponding
to a given choice of both tensor product structure and 
coefficients $\{a_{ij}\}$ 
described above, corresponds to the class of systems like (13) but
with undetermined coefficients $Q^u$ and has the same form
\begin{equation}
L^u(Q^u)\Psi^u=0.
\end{equation}
Our goal is to restore coefficients $Q^u$ from (25), (26) and 
explicit representations for solutions 
$\Psi^0$ and $\Psi^u$. This is a standard problem in the adaptive control
theory \cite{11}: one adds a controlling signal
$u(x,t)$ which deforms the controlled signal $\Psi(x,t)$ from the fixed state
$\Psi^0(x,t)$ to the prescribed one $\Psi^u(x,t)$. At the same time one can 
determine the 
 parameters $Q^u$ \cite{3}.
Finally, we apply two variational constructions.
The first one gives the systems of algebraic equations for  
unknown coefficients, generated by the following set of functionals
\begin{equation}
\Phi_N=\int\Big((L^0-L^u)\Psi^u_N,\Psi^0_N\Big){\rm d}\mu_N,
\end{equation}
where $N$ means the $N$-order approximation according to formulae (15). 
The unknown parameters $Q^*$ are given by $Q^*=\lim_{N\to\infty}Q^u_N$.
The second is an important additional constraint on the region 
$\mu_0$ in the phase space 
where we are interested in localization of almost all energy
$E=\int H(\Psi^u){\rm d}\mu$,
where $E$ is the proper energy functional \cite{2} (Marsden-like). 

We believe that the appearance of nontrivial localized patterns 
observed by these simple methods is a general effect which is also present in the 
full BBGKY--hierarchy, due to its complicated 
intrinsic multiscale dynamics 
and it depends on neither the cut-off level nor the phenomenological-like
hypothesis on correlators. So, representations like (17) and the prediction of the existence of
the (asymptotically) stable localized patterns/states (energy confinement states) 
in BBGKY-like systems are 
the main results of this paper.

\section*{Acknowledgments}  
We are very grateful to M. Bonitz for invaluable discussions, encouragement and support.

\end{document}